\documentclass[10pt, a4paper]{article}
\pdfoutput=1

\usepackage[utf8]{inputenc}
\usepackage[T1]{fontenc}
\usepackage{lmodern,textcomp}
\usepackage[english]{babel}
\usepackage{csquotes}

\usepackage[top=4cm,bottom=4cm,left=3.5cm,right=3.5cm]{geometry}

\usepackage{eurosym,xspace}
\usepackage[nottoc]{tocbibind}
\usepackage{graphicx,subfig,float}
\usepackage[multiple]{footmisc}
\usepackage{authblk}
\usepackage[backend=bibtex8, citestyle=numeric-comp, maxbibnames=99, sorting=nyt,
			sortcites=true, firstinits=true, sorting=none]{biblatex}
\input{bibphys.def}

\usepackage{amsmath,amsfonts,amssymb}

\usepackage{maths_min}

\usepackage[pdftex, colorlinks=true, breaklinks=true,
	urlcolor=blue, linkcolor=blue, citecolor=red,
	pdfcreator={PDFLaTeX}, pdfauthor={Harold Erbin}
]{hyperref}
\usepackage[all]{hypcap}

\hypersetup{
	pdftitle={Janis-Newman algorithm: simplifications and gauge field transformation},
}

\numberwithin{equation}{section}

\newcommand{\email}[1]{\thanks{\href{mailto:#1}{\texttt{#1}}}}

\title{Janis-Newman algorithm: simplifications and gauge field transformation}

\author[1]{Harold Erbin\email{erbin@lpthe.jussieu.fr}}
\affil[1]{Sorbonne Universités, UPMC Univ Paris 06, UMR 7589, LPTHE, F-75005, Paris, France}
\affil[1]{CNRS, UMR 7589, LPTHE, F-75005, Paris, France}

\begin{document}

\maketitle

\begin{abstract}
The Janis–Newman algorithm is an old but very powerful tool to generate rotating solutions from static ones through a set of complex coordinate transformations.
Several solutions have been derived in this way, including solutions with gauge fields.
However, the transformation of the latter was so far always postulated as an ad hoc result.
In this paper we propose a generalization of the procedure, extending it to the transformation of the gauge field.
We also present a simplification of the algorithm due to G. Giampieri.
We illustrate our prescription on the Kerr–Newman solution.
\end{abstract}

\tableofcontents
\newpage

\section{Introduction}

Originally Janis and Newman presented a "derivation" of the Kerr metric using a complex coordinate transformation~\cite{newman_note_1965}.
Although there was at that time no clear reason why this should work, several groups started to use it, discovering new axisymmetric solutions of Einstein equations, possibly coupled to Maxwell~\cite{newman_metric_1965} or dilaton~\cite{yazadjiev_newman-janis_2000} fields, with a cosmological constant~\cite{demianski_new_1972}, and later to interior solutions~\cite{herrera_complexification_1982, drake_application_1997, drake_explanation_2000, ibohal_rotating_2005} and other dimensions~\cite{xu_exact_1988, kim_notes_1997, kim_spinning_1999}.

The so called Janis–Newman (JN) algorithm provides a way to generate axisymmetric metrics from a (spherical) seed metric through a particular complexification of radial and time coordinates, followed by a complex coordinate transformation.
Despite the fact that the complexification relies mostly on intuitive guesses and that there exists no strong generic formulation for it, some general features have been worked out in the last decades.
For instance, it has been shown that the complexification is unique~\cite{drake_explanation_2000} under some assumptions, and – as we will see – that some different complexifications are equivalent.
Detailed reviews on generalizations and explanations of the JN algorithm can be found in~\cites{drake_explanation_2000}[sec.~5.4]{whisker_braneworld_2008} while justifications may be found in~\cite{talbot_newman-penrose_1969, schiffer_kerr_1973, gurses_lorentz_1975, ferraro_untangling_2014}.

The JN procedure can appear to be very tedious since it requires to invert the metric, to find a null tetrad basis where the transformation can be applied, and lastly to invert again the metric.
In~\cite{giampieri_introducing_1990} Giampieri introduced another formulation of the JN algorithm which avoids gymnastics with null tetrads and which appears to be very useful for extending the procedure to more complicated solutions (such as higher dimensional ones).
However it has been so far totally ignored in the literature and we wish to bring attention on it.

As far as gauge fields are concerned, Newman et al. could not obtain the electromagnetic field strength from the original version of the Janis–Newman algorithm~\cite{newman_metric_1965}.
As a consequence all the papers dealing with solutions including a gauge field are making an ansatz for the latter, which feels unsatisfactory.
Performing a suitable gauge transformation to remove the radial component of the gauge field is the key for applying the algorithm with either prescription.
As for the metric, the JN approach needs to be carried out on the contravariant components and accordingly Giampieri's prescription is simpler.

It is worth mentioning that another solution was recently proposed in~\cite{keane_extension_2014}, where a null Lorentz transformation on the tetrads is used to obtain the correct Newman–Penrose coefficients for the field strength.
According to our proposal, we claim that it is possible to perform such transformations with a much more natural formulation, using the gauge field being more convenient than using the field strength or its Newman–Penrose coefficients (for example in view of matter coupling).
Our approach is also closer to the original spirit of the algorithm as one works with contravariant components (written with tetrads) for both the metric and the gauge field, and the transformation follows the same pattern.

The paper is organized as follows.
In section~\ref{sec:original-jna} we outline the main steps of the JN algorithm as it was prescribed in its original version~\cite{newman_note_1965, newman_metric_1965}.
We expose in section~\ref{sec:giampieri-jna} Giampieri's \emph{effective} prescription of the JN algorithm~\cite{giampieri_introducing_1990}.
In the section~\ref{sec:kerr-newman} we show how to transform the gauge field, exhibiting the procedure on the original JN example which is the Kerr–Newman black hole.
Furthermore our prescription can be very useful for numerical calculations since all transformations can actually be applied at the same time, as described in appendix~\ref{sec:chaining-transformations}.
In appendix~\ref{sec:arbitrariness} we discuss the arbitrariness of the complexification.

\section{Janis–Newman algorithm}
\label{sec:original-jna}

In their original paper~\cite{newman_note_1965}, Janis and Newman demonstrated how to recover the Kerr metric from the Schwarzschild one.
In this section we outline the procedure with the seed metric
\begin{equation}
	\label{metric:spherical:f}
	\dd s^2 = - f(r)\, \dd t^2 + f(r)^{-1}\, \dd r^2 + r^2 \dd \Omega^2, \qquad
	\dd \Omega^2 = \dd\theta^2 + \sin^2 \theta\; \dd \phi^2.
\end{equation}

The algorithm proceeds as follows:
\begin{enumerate}
	\item Introduce the null coordinate
	\begin{equation}
		\label{change:tr-ur}
		\dd u = \dd t - f^{-1} \dd r \,.
	\end{equation} 
	The metric becomes
	\begin{equation}
		\dd s^2 = - f\, \dd u^2 - 2\, \dd u \dd r + \dd \Omega^2.
	\end{equation} 
	
	\item Find the contravariant form of the metric $g^{\mu\nu}$ and use the null tetrad $\{\ell^\mu, n^\mu, m^\mu, \bar m^\mu\}$ to express it as
	\begin{equation}
		g^{\mu\nu} = - \ell^\mu n^\nu - \ell^\nu n^\mu + m^\mu \bar m^\nu + m^\nu \bar m^\mu,
	\end{equation} 
	where the vectors are taken to be
	\begin{equation}
		\label{eq:tetrads:static}
		\ell^\mu = \delta_r^\mu, \qquad
		n^\mu = \delta_u^\mu -\frac{f}{2}\; \delta_r^\mu, \qquad
		m^\mu = \frac{1}{\sqrt{2} \bar r} \left(\delta_\theta^\mu + \frac{i}{\sin \theta} \delta_\phi^\mu \right).
	\end{equation} 
	At this point $r$ is real such that $\bar r = r$.
	
	\item Allow the coordinates $u$ and $r$ to take complex values together with the conditions:
	\begin{itemize}
		\item $\ell^\mu$ and $n^\mu$ must be kept real;
		\item $m^\mu$ and $\bar m^\mu$ must still be complex conjugated to each other;
		\item one should recover the previous basis for $u, r \in \R$.
	\end{itemize}
	The previous conditions imply that the function $f(r)$ should be replaced by a new function $\tilde f(r, \bar r) \in \R$ such that $\tilde f(r, r) = f(r)$.
	This step is the hardest to perform because there is no \emph{a priori} rule to choose any particular complexification and one needs to check systematically if Einstein equations are satisfied.
	Examples have provided a set of rules that can be used
	\begin{subequations}
	\label{eq:complexification-rules}
	\begin{align}
		\label{eq:complexification-rules-r}
		r & \longrightarrow \frac{1}{2} (r + \bar r) = \Re r\,, \\
		\label{eq:complexification-rules-1/r}
		\frac{1}{r} & \longrightarrow \frac{1}{2} \left(\frac{1}{r} + \frac{1}{\bar r}\right) = \frac{\Re r}{\abs{r}^2}\,, \\
		\label{eq:complexification-rules-r2}
		r^2 & \longrightarrow \abs{r}^2.
	\end{align}
	\end{subequations}
	
	\item Carry out a complex change of coordinates~\footnote{This transformation can be made more general~\cite{drake_explanation_2000, demianski_new_1972, whisker_braneworld_2008}.}
	\begin{equation}
		\label{change:complexification-ur}
		u = u' + ia \cos \theta, \qquad
		r = r' - ia \cos \theta, \qquad
		\theta' = \theta, \qquad
		\phi' = \phi,
	\end{equation} 
	$a$ being a parameter (with the interpretation of angular momentum per unit of mass), with the restriction that $r', u' \in \R$.
	The tetrads transform as vectors and now $\tilde f = \tilde f(r', \theta')$ (but note that the $\theta'$ dependence is not arbitrary and comes solely from $\Im z$).
	
	Explicitly one gets (forgetting the primes for convenience)
	\begin{equation}
	\begin{gathered}
		\label{eq:tetrads:rotating}
		\ell'^\mu = \delta_r^\mu, \qquad
		n'^\mu = \delta_u^\mu - \frac{\tilde f}{2}\; \delta_r^\mu, \\
		m'^\mu = \frac{1}{\sqrt{2} (r + ia \cos \theta)} \left(\delta_\theta^\mu + \frac{i}{\sin \theta} \delta_\phi^\mu - ia \sin \theta\, (\delta_u^\mu - \delta_r^\mu) \right).
	\end{gathered}
	\end{equation} 
	
	\item Construct the metric $g^{\mu\nu}$ from the new set of tetrads and invert it.
	
	\item Eventually change the coordinates into any other preferred system, e.g. Boyer–Lind\-quist. If the transformation is infinitesimal then one should check that it is a valid diffeomorphism, i.e. that it is integrable.
\end{enumerate}

\section{Giampieri's formulation}
\label{sec:giampieri-jna}

In the former approach it is very tedious to invert twice the metric and find out the right tetrad basis.
In an essay submitted to the \emph{Gravity Research Foundation}~\cite{giampieri_introducing_1990}, Giampieri proposed a simplification to this algorithm.
In a nutshell, coordinates $u$ and $r$ are complexified in the metric itself and we change coordinates directly in the metric.
Then all complex $i$ factors are removed using a specific ansatz for the coordinate transformation.

Giampieri applied his method only to the Schwarzschild metric, thus it is worth to detail it in the more general context of \eqref{metric:spherical:f} with arbitrary $f$.
The procedure is the following:
\begin{enumerate}
	\item Introduce the null coordinate $u$
	\begin{equation}
		\label{metric:spherical:null}
		\dd s^2 = -f\, \dd u^2 - 2\, \dd u \dd r + \dd \Omega^2.
	\end{equation} 
	
	\item Allow the coordinates $u$ and $r$ to take complex values and complexify the metric \eqref{metric:spherical:null} to
	\begin{equation}
		\dd s'^2 = - \tilde f\, \dd u^2 - 2\, \dd u \dd r + \abs{r}^2 \dd \Omega^2,
	\end{equation} 
	using the rules \eqref{eq:complexification-rules-r2} for the coefficient of $\dd\Omega^2$ and where again $\tilde f = \tilde f(r, \bar r)$ is the real-valued function which is replacing $f$.
	At this step the metric continues being real.
	
	\item Apply the change of coordinates \eqref{change:complexification-ur}
	\begin{equation}
		\label{change:jna-ur}
		u = u' + ia \cos \psi, \qquad
		r = r' - ia \cos \psi, \qquad
		\theta' = \theta, \qquad
		\phi' = \phi,
	\end{equation} 
	where a new angle $\psi$ is introduced.
	This amounts to embedding the spacetime in a $5$-dimensional complex spacetime and the final metric will correspond to a $4$-dimensional real slice.
	The differentials read
	\begin{equation}
		\label{change:jna-diff-ur-old}
		\dd u = \dd u' - ia \sin \psi\; \dd \psi, \qquad
		\dd r = \dd r' + ia \sin \psi\; \dd \psi,
	\end{equation}
	and one gets the metric
	\begin{equation}
		\begin{aligned}
			\dd s'^2  = - \tilde f (&\dd u - ia \sin\psi\, \dd\psi)^2
				- 2\, (\dd u - ia \sin\psi\, \dd\psi) (\dd r + ia \sin\psi\, \dd\psi) \\
				&+ (r^2 + a^2 \cos^2 \theta)\, \dd \Omega^2.
		\end{aligned}
	\end{equation}
	
	\item As one can easily notice, this metric can not be correct because it has to be real.
	Giampieri found that this metric reduces to the result from the original formulation if one uses the ansatz
	\begin{subequations}
	\label{eq:giampieri-ansatz}
	\begin{equation}
		i\, \dd \psi = \sin \psi\, \dd \phi
	\end{equation}
	followed by the replacement
	\begin{equation}
		\psi = \theta.
	\end{equation}
	\end{subequations}
	This step is absolutely ad hoc but, as we will see, leads nicely to the right solution. The question whether it can be generalized to a larger class of solution is addressed in a separate work and will be the object of a following paper.
	
	Deleting all the primes, the metric obtained in the Kerr coordinates~\cite{newman_note_1965} is
	\begin{equation}
		\label{metric:rotating:Kerr-coord}
		\dd s^2  = - \tilde f\, (\dd u - a \sin^2 \theta\, \dd\phi)^2
			- 2\, (\dd u - a \sin^2 \theta\, \dd\phi) (\dd r + a \sin^2 \theta\, \dd\phi)
			+ \rho^2 \dd \Omega^2
	\end{equation}
	where we have introduced
	\begin{equation}
		\label{metric:rotating:rho}
		\rho^2 = r^2 + a^2 \cos^2 \theta.
	\end{equation}
	
	\item Finally one can go to Boyer–Lindquist coordinates with
	\begin{equation}
		\label{change:rotating:g-h}
		\dd u = \dd t' - g(r) \dd r, \qquad
		\dd \phi = \dd \phi' - h(r) \dd r.
	\end{equation} 
	
	The conditions $g_{tr} = g_{r\phi'} = 0$ are solved for
	\begin{equation}
		\label{change:rotating:ur-bl}
		g = \frac{r^2 + a^2}{\Delta}, \qquad
		h = \frac{a}{\Delta}
	\end{equation} 
	where we have defined
	\begin{equation}
		\label{metric:rotating:delta}
		\Delta = \tilde f \rho^2 + a^2 \sin^2 \theta.
	\end{equation} 
	As indicated by the $r$-dependence this change of variable is integrable provided that $g$ and $h$ are functions of $r$ only.
	However $\Delta$ as given in \eqref{metric:rotating:delta} could in principle contain a dependence on $\theta$, thus it is absolutely essential that one checks that this is not the case.
	Similarly the complex transformation does not preserve Einstein equations in general and they need to be verified before claiming that a new solutions has been found.
	These two points are of particular importance since several metrics derived from JN algorithm~\cites[sec.~5.4.2]{whisker_braneworld_2008}{mallett_metric_1988, capozziello_axially_2010, caravelli_spinning_2010, ghosh_radiating_2013} have been shown to be wrong for one of these reasons~\cite{xu_radiating_1998, azreg-ainou_comment_2011, azreg-ainou_generating_2014}.
	
	Given this condition one gets the metric (deleting the prime)~\cite[p.~14]{visser_kerr_2007}
	\begin{equation}
		\label{metric:rotating:bl-coord}
		\dd s^2 = - \tilde f\, \dd t^2
			+ \frac{\rho^2}{\Delta}\, \dd r^2
			+ \rho^2 \dd\theta^2
			+ \frac{\Sigma^2}{\rho^2} \sin^2 \theta\; \dd\phi^2
			+ 2a (\tilde f - 1) \sin^2 \theta\; \dd t \dd\phi
	\end{equation} 
	with
	\begin{equation}
		\frac{\Sigma^2}{\rho^2} = r^2 + a^2 + a g_{t\phi} \,.
	\end{equation} 
	
\end{enumerate}

We stress that the order of the steps should be respected, otherwise the ansatz \eqref{eq:giampieri-ansatz} can not be consistently applied.
The second important point is that JN and Giampieri's prescriptions differ only in the computation of the metric since the rules \eqref{eq:complexification-rules} are identical in both cases.
Therefore this new approach is not adding or removing any of the ambiguity that is already present and well-known in JN algorithm.
In particular the ansatz \eqref{eq:giampieri-ansatz} is a direct consequence of the fact that the $2$-dimensional slice $(\theta, \phi)$ is given by
\begin{equation}
	\dd \Omega^2 = \dd\theta^2 + \sin^2 \theta\; \dd \phi^2,
\end{equation} 
the function in the RHS of \eqref{eq:giampieri-ansatz} corresponding to $\sqrt{g_{\phi\phi}}$ (where $g$ is the static metric) as can be seen by doing the computation with $i\, \dd \psi = H(\psi) \dd \phi$ and identifying $H$ at the end.

Comparing \eqref{metric:spherical:null} and \eqref{metric:rotating:Kerr-coord} makes clear that the effect of the ansatz \eqref{eq:giampieri-ansatz} can be reduced to modifying the formula \eqref{change:jna-diff-ur-old} into
\begin{equation}
	\label{change:jna-diff-ur}
	\dd u = \dd u' - a \sin^2 \theta\; \dd \phi\,, \qquad
	\dd r = \dd r' + a \sin^2 \theta\; \dd \phi\,.
\end{equation}

Using directly these expressions allows to avoid introducing the angle $\psi$ altogether.
Although some authors~\cite{ibohal_rotating_2005, ferraro_untangling_2014} mentioned the equivalence of these formulae and the result from the tetrads as a curiosity, it is surprising that this direction has not been followed further.

\section{Kerr–Newman: transforming the gauge field}
\label{sec:kerr-newman}

In this section we apply the formalism to the Reissner–Nordström black hole in order to get the Kerr–Newman rotating black hole, both of which are solutions of Einstein–Maxwell theory.
The metric is obtained first while the gauge field is found using both JN and Giampieri's prescriptions.

The seed solution corresponds to the metric
\begin{equation}
	\label{metric:reissner-nordstrom:tr}
	\dd s^2 = - f(r)\, \dd t^2 + f(r)^{-1}\, \dd r^2 + r^2 \dd \Omega^2, \qquad
	f(r) = 1 - \frac{2m}{r} + \frac{q^2}{r^2}
\end{equation} 
and to the gauge field
\begin{equation}
	\label{pot:reissner-nordstrom:tr}
	A = \frac{q}{r}\; \dd t
\end{equation} 
where the parameters $m$ and $q$ correspond respectively to the mass and to the electric charge.

Using the rules \eqref{eq:complexification-rules-1/r} and \eqref{eq:complexification-rules-r2} for the second and third terms respectively, the function $f$ can be complexified as
\begin{equation}
	\label{eq:kerr-newman:f}
	\tilde f(r, \theta) = 1 + \frac{q^2 - 2m r}{\rho^2}
\end{equation}
where we recall that $\rho^2 = \abs{r}^2 = r^2 + a^2 \cos^2 \theta$.

As already described in~\cite{drake_explanation_2000, newman_metric_1965}, plugging this function into \eqref{metric:rotating:bl-coord} gives the well-known Kerr–Newman metric
\begin{equation}
	\dd s^2 = - \tilde f\, \dd t^2
		+ \frac{\rho^2}{\Delta} \dd r^2
		+ \rho^2 \dd\theta^2
		+ \frac{\Sigma^2}{\rho^2} \sin^2 \theta\; \dd\phi^2
		+ 2a (\tilde f - 1) \sin^2 \theta\; \dd t \dd\phi\,,
\end{equation}
where functions $\Delta$ and $\Sigma$ are given by
\begin{subequations}
\begin{gather}
	\frac{\Sigma^2}{\rho^2} = r^2 + a^2 - \frac{q^2 - 2Mr}{\rho^2}\, a^2 \sin^2 \theta \,, \\
	\Delta = r^2 - 2Mr + a^2 + q^2,
\end{gather}
\end{subequations}
and it is to point out that $\Delta$ depends only on $r$ so that the transformation \eqref{change:rotating:ur-bl} to Boyer–Lindquist coordinates is well defined.

\subsection{Giampieri's formalism}

As already mentioned in the introduction, the authors of~\cite{newman_metric_1965} face serious difficulties while trying to derive the field strength of the Kerr–Newman black hole from the Reissner–Nordström one.
Indeed, in the null tetrad formalism, the field strength is given in terms of Newman\-Penrose coefficients and problems arise when trying to generate the rotating solution since one of the coefficients, vanishing in the case of Reissner–Nordström, is non-zero for Kerr–New\-man.
We show that using Giampieri's prescription allows to circumvent the problem in a very simple way.

Starting with the gauge field \eqref{pot:reissner-nordstrom:tr} for the Reissner–Nordström black hole and expressing it in terms of the $(u, r)$ coordinates gives
\begin{equation}
	A = \frac{q}{r}\, (\dd u + f^{-1} \dd r).
\end{equation}  
The second term actually does not contribute to the field strength since $A_r = A_r(r)$ and one can remove it by a gauge transformation, getting
\begin{equation}
	\label{pot:reissner-nordstrom:ur}
	A = \frac{q}{r}\; \dd u \,.
\end{equation} 

Applying the transformations \eqref{change:jna-diff-ur} gives
\begin{equation}
	A' = \frac{q r}{\rho^2}\, (\dd u - a \sin^2 \theta\, \dd \phi)
\end{equation} 
where as usual $\rho^2 = r^2 + a^2 \cos^2 \theta$.
The prefactor here has been transformed using the rule \eqref{eq:complexification-rules-1/r}.

Going to Boyer–Lindquist coordinates, using \eqref{change:rotating:ur-bl}, provides
\begin{equation}
	A' = \frac{q r}{\rho^2}\, \left(\dd u - \frac{\rho^2}{\Delta}\, \dd r - a \sin^2 \theta\, \dd \phi \right).
\end{equation} 

Finally, the factor $\rho^2$ in front of $\dd r$ cancels with the prefactor, and we are left with
\begin{equation}
	A'_r = \frac{q r}{\Delta}
\end{equation} 
which depends only on $r$.
This can again be removed by a gauge transformation, and one obtains the traditional form of the electromagnetic gauge field for the Kerr–Newman black hole (omitting the prime)
\begin{equation}
	A = \frac{q r}{\rho^2}\, (\dd t - a \sin^2 \theta\, \dd \phi).
\end{equation} 

\subsection{Tetrad formalism}

Expression \eqref{pot:reissner-nordstrom:ur} for the static gauge potential – after the gauge transformation – can be rewritten as
\begin{equation}
	A_\mu = \frac{q}{r}\, \delta_\mu^u.
\end{equation} 
Using the inverse of the metric \eqref{metric:spherical:null} with function \eqref{metric:reissner-nordstrom:tr} one obtains the contravariant expression
\begin{equation}
	A^\mu = - \frac{q}{r}\, \delta_r^\mu = - \frac{q}{r}\, \ell^\mu
\end{equation} 
where $\ell^\mu = \delta_r^\mu$, see \eqref{eq:tetrads:static}.

The JN transformation applied to the previous expression yields
\begin{equation}
	A'^\mu = - \frac{q r}{\rho^2}\, \ell'^\mu = - \frac{q r}{\rho^2}\, \delta_r^\mu
\end{equation} 
with $\ell'^\mu = \ell^\mu$ is defined in \eqref{eq:tetrads:rotating}.
Finally the $1$-form
\begin{equation}
	A' = \frac{q r}{\rho^2}\, (\dd u - a \sin^2 \theta\, \dd \phi)
\end{equation} 
is retrieved using the metric \eqref{metric:rotating:Kerr-coord} with the function \eqref{eq:kerr-newman:f}.

The result is identical to the one derived with Giampieri's formalism, showing again that the two approaches are totally equivalent, and that it was not necessary to use the null Lorentz rotation from~\cite{keane_extension_2014}.
It is possible to check that the transformation can not be performed without first removing the $r$-component with the gauge transformation.

\section{Conclusion}
\label{sec:conclusion}

As it was announced in the introduction, this paper contains two main ideas.
First the Giamperi's prescription for performing the Janis–Newman algorithm turns out to be a tool which deserves to be better understood and generalized to other classes of solutions.
This is the aim of different works in progress.

The second conclusion of this paper is that just as rotating metrics can be derived from static ones through the Giamperi's and tetrad formalisms, the gauge field can be automatically derived as well.
As far as we know, this idea is completely new and opens the door to many possible extensions of Janis–Newman original idea.

\section*{Acknowledgments}

I would like to thank Lucien Heurtier for many discussions, collaborations on related topics and for corrections on the draft. I am also very grateful to Tresa Bautista and Eric Huguet for having carefully read and commented the draft, and to Nick Halmagyi for interesting discussions and for bringing the original JN papers~\cite{newman_note_1965, newman_metric_1965} to my attention.

This work, made within the \textsc{Labex Ilp} (reference \textsc{Anr–10–Labx–63}), was supported by French state funds managed by the \emph{Agence nationale de la recherche}, as part of the programme \emph{Investissements d'avenir} under the reference \textsc{Anr–11–Idex–0004–02}.

\appendix

\section{Chaining transformations}
\label{sec:chaining-transformations}

The JN algorithm is summarized by the following table
\begin{equation}
	\begin{array}{cccccccccc}
 		t             & \to & u & \to & u \in \C    & \to & u' & \to & t'    \\
		r             &     &   & \to & r \in \C    & \to & r' &     &      \\
		\phi          &     &   &     &             &     &    & \to & \phi' \\
		f             &     &   & \to & \tilde f    &     &    &     &      \\
		g_{\mu\nu}    &     &   & \to & g'_{\mu\nu} &     &    &     &
	\end{array}
\end{equation}
where the arrows correspond respectively to the steps 1, 2, 4 and 5 of section~\ref{sec:giampieri-jna} (and 1, 3, 4 and 6 of section~\ref{sec:original-jna}).

A major advantage of Giampieri's prescription is that one can chain all these transformations since it involves only substitutions and no tensor operations.
For this reason it is much easier to implement on a computer algebra system such as Mathematica.
It is then possible to perform a unique change of variables that leads directly from the static metric to the rotating metric in any system defined by the function $(g, h)$
\begin{subequations}
\begin{align}
	\dd t &= \dd t' + \big(a h \sin^2 \theta\, (1 - \tilde f^{-1}) - g + \tilde f^{-1} \big)\, \dd r'
		+ a \sin^2 \theta\, (\tilde f^{-1} - 1)\, \dd \phi', \\
	\dd r &= (1 - a h \sin^2 \theta)\, \dd r' + a \sin^2 \theta\; \dd \phi', \\
	\dd \phi &= \dd \phi' - h\, \dd r',
\end{align}
\end{subequations}
where the complexification of the metric function $f$ can be made at the end.
It is impressive that steps 1 to 5 from section~\ref{sec:giampieri-jna} can be written in such a compact way.

\section{Arbitrariness of the complexification}
\label{sec:arbitrariness}

In this appendix we provide a short comment on the arbitrariness of the complexification rules.
In particular let's consider the functions
\begin{equation}
	f_1(r) = \frac{1}{r}, \qquad
	f_2(r) = \frac{1}{r^2}.
\end{equation} 

The usual rule is to complexify these two functions as
\begin{equation}
	\label{eq:arb-usual-rules}
	\tilde f_1(r) = \frac{\Re r}{\abs{r}^2}, \qquad
	\tilde f_2(r) = \frac{1}{\abs{r}^2}
\end{equation} 
using respectively the rules \eqref{eq:complexification-rules-1/r} and \eqref{eq:complexification-rules-r2} (in the denominator).

But it is possible to arrive at the same result with a different combinations of rules.
In fact the functions can be rewritten as
\begin{equation}
	f_1(r) = \frac{r}{r^2}, \qquad
	f_2(r) = \frac{1}{r}\, \frac{1}{r}.
\end{equation} 
The following set of rules results again in \eqref{eq:arb-usual-rules}:
\begin{itemize}
	\item $f_1$: \eqref{eq:complexification-rules-r} (numerator) and \eqref{eq:complexification-rules-r2} (denominator);
	\item $f_2$: \eqref{eq:complexification-rules-r} (first fraction) and \eqref{eq:complexification-rules-1/r} (second fraction).
\end{itemize}

\printbibliography[heading=bibintoc]

\end{document}